\documentclass[aps,prl,twocolumn,showpacs,preprintnumbers,groupedaddress]{revtex4}
\usepackage{graphicx}

\begin{document}

\title{
Observation of large many-body Coulomb interaction effects in a doped quantum wire
}

\author{
Hidefumi Akiyama${}^{1,2,3}$, Loren N. Pfeiffer${}^1$, Aron Pinczuk${}^{1,2}$, Ken W. West${}^1$, and Masahiro Yoshita${}^3$
}

\affiliation{
${}^1$ Bell Laboratories, Lucent Technologies, 600 Mountain Avenue, Murray Hill, NJ 07974, USA, \\
${}^2$ Department of Physics, Columbia University, New York, NY 10027, USA, \\
${}^3$ Institute for Solid State Physics, University of Tokyo, 5-1-5 Kashiwanoha, Kashiwa, Chiba 277-8581, Japan\\ }

\date{\today}

\begin{abstract}
We demonstrate strong one dimensional (1-D) many-body interaction effects in photoluminescence (PL) in a GaAs single quantum wire of unprecedented optical quality, where 1-D electron plasma densities are controlled via electrical gating. We observed PL of 1-D charged excitons with large binding energy of 2.3 meV relative to the neutral excitons, and its evolution to a Fermi-edge singularity at high electron density. Furthermore, we find a strong band-gap renormalization in the 1-D wire, or a large red-shift of PL with increased electron plasma density. Such a large PL red-shift is not observed when we create a high density neutral electron-hole plasma in the same wire, due probably to cancellation of the Coulomb interaction energy in the neutral plasma. 
\end{abstract}

\maketitle


Many-body photoluminescence (PL) effects in semiconductor quantum wires (QWRs) with electron gases are expected to become strong or anomalous due to singular nature of the Coulomb interaction in one-dimensional (1-D) systems \cite{OgawaPRL,BennerEL}. Shrinkage of the optical band gap with an electron plasma due to the many-body Coulomb interaction is called band-gap renormalization (BGR), which causes well-known red-shifts of PL spectra in 2-D quantum wells (QWs). The possible existence of strong 1-D Coulomb interactions and their implications for BGR effects inherent to 1-D systems has been an issue of current fundamental interest, and indeed intensive theoretical studies have been made \cite{HuPRL,RossiPRB,TassonePRL,DassarmaPRL}. 

Experimentally, however, there are only few reports on PL of doped single-mode QWRs \cite{RinaldiPRB,CallejaSSC,JKimPE}, and of these one claimed reduced BGR effects in doped QWRs \cite{RinaldiPRB}. 
In highly-photo-excited undoped QWRs \cite{WegscheiderPRL,AmbigapathyPRL,SiriguPE,TGKimPE}, lasing and PL experiments have shown no BGR effect.
Because of technical limitations in carrier density control, reproducibility, and PL line broadening due to structural inhomogeneities of the QWRs, consistency among the various observations and with theory has not yet been established for 1-D QWRs. 

In this Letter, we report on PL experiments in an n-type-modulation-doped single GaAs QWR of superior quality, in which we tune the electron density with application of an external electric field. The interplay between the 1-D excitonic interaction and BGR is studied in a single QWR sample, where in addition to the neutral exciton {\it we observed a clearly resolved 1D charged exciton, large BGR in a 1-D electron plasma, and substantially smaller BGR in a 1-D electron-hole plasma}:  PL spectra at moderate electron densities are dominated by the 1-D charged excitons, that are significantly shifted from the neutral excitons. At high electron densities above $2.5\times10^5$ cm${}^{-1}$, the PL spectra exhibit striking lineshapes in which the onsets (band edge) of emission continuously shift to the red with increasing electron density and the high-energy cutoff (Fermi edge) of emission is pinned at the position of the charged exciton measured at small electron density. The large red-shifts in the PL reveal the major impact of BGR on optical recombination in the 1-D electron plasma. In contrast, we find much smaller red-shift in PL for neutral two-component electron-hole plasmas, when we gate off the electrons in the same QWR and control the neutral electron-hole plasma by the intensity of optical pumping. 
These results suggest that Coulomb interaction effects are significant in QWRs and that the intriguing difference between electron and electron-hole plasmas result from the first order cancellation of the Coulomb interactions in an electron-hole plasma with charge neutrality. 

\begin{figure}
\includegraphics[width=.4\textwidth]{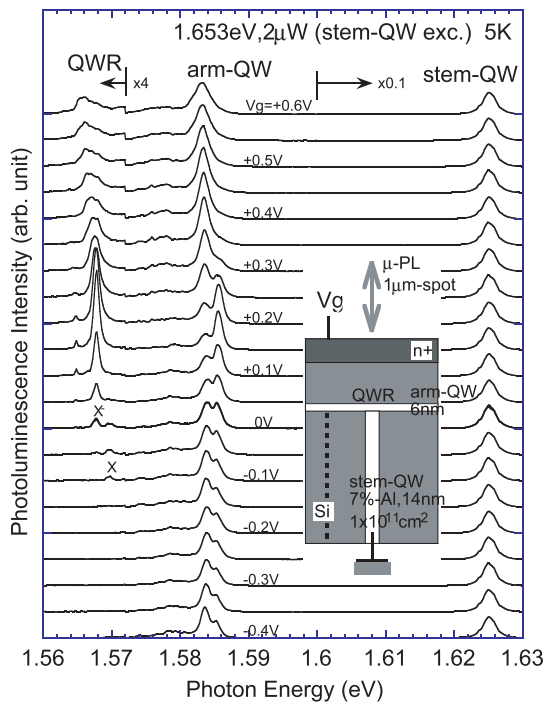}
\caption{
PL spectra at 5 K for various gate voltages $V_g$ in 0.05 V steps. The inset shows the schematic structure of T-shaped QWR, stem-QW, arm-QW, and a gate. Electrons are accumulated or depleted in the QWR by positive or negative $V_g$. 
}
\label{1}
\end{figure}

The sample was grown by the cleaved-edge overgrowth (CEO) method with molecular beam epitaxy (MBE) \cite{PfeifferAPL}.
First, on a non-doped (001) GaAs substrate we successively grew a 50 nm GaAs buffer layer, a 2.26 $\mu$m barrier layer of (GaAs)${}_9$(Al${}_{0.33}$Ga${}_{0.67}$As) ${}_{71}$ super-lattice, a 14 nm Al${}_{0.07}$Ga${}_{0.93}$As QW (stem-QW) layer, a 100 nm Al${}_{0.33}$Ga${}_{0.67}$As spacer layer, a 4$\times10^{11}$ cm${}^{-2}$ Si delta-doping layer, a 100 nm Al${}_{0.33}$Ga${}_{0.67}$As barrier layer, a 5.66 $\mu$m barrier layer of (GaAs)${}_9$(Al${}_{0.33}$Ga${}_{0.67}$As) ${}_{71}$ super-lattice, and a 30 nm GaAs cap layer. Then, in a separate MBE growth on an {\it in-situ} cleaved (110) edge of this structure, we grew at 490 ${}^{\circ}$C by the CEO method, a 6 nm GaAs QW (arm-QW) layer, a 200 nm Al${}_{0.45}$Ga${}_{0.55}$As barrier layer, and a 100nm heavily-Si-doped n${}^+$-Al${}_{0.1}$Ga${}_{0.9}$As layer. Right after the growth of the GaAs arm-QW layer, growth was interrupted for 10 minute anneal at 600 ${}^{\circ}$C, which is based on our recently-developed growth-interruption annealing technique \cite{YoshitaAPL} that dramatically reduces monolayer fluctuation on a (110)-GaAs surface \cite{YoshitaPRB}.

As shown in the inset of Fig. 1, a single QWR is formed at the T-shaped intersection of the stem-QW and the arm-QW.  By applying positive or negative DC gate voltage $V_g$ to the n${}^+$-Al${}_{0.1}$Ga${}_{0.9}$As layer relative to modulation-doped 2-D electron gas in the stem-QW with density of $1\times10^{11}$ cm${}^{-2}$, we accumulated or depleted additional electrons in the QWR. 
Micro-PL measurements on the single QWR were performed with a cw titanium-sapphire laser via the (110) surface in the geometry  of point excitation into a 0.8 $\mu$m spot \cite{YoshitaJAP}. A 0.6m triple spectrometer and a back-illumination-type liquid-nitrogen-cooled CCD camera were used to detect the PL. 

Figure 1 shows PL spectra at 5 K for various gate voltages $V_g$ in 0.05 V steps between $-0.4$ and 0.6 V. Spectral resolution was 0.3 meV. The excitation light with intensity of 2 $\mu$W and photon energy of 1.653 eV was incident normal to the (110) surface, and initially created carriers mainly in the stem-QWs, some portion of which flowed into the QWR. 
Well-resolved sharp peaks were observed for the QWR (1.568 eV), the arm-QW (1.584 and 1.586 eV), and the stem-QW (1.625 eV) in each $V_g$ gated spectrum. 
The low energy PL tail of the arm-QW around 1.572-1.582 eV shows localized states in the arm-QW. 
However, the sharpness of the QWR and arm-QW peaks with full-width-of-half-maximum (FWHM) of about 1 meV is about an order of magnitude smaller than those reported in all previous work \cite{AkiyamaJP}, and reveals the high flatness of the (110) interfaces.

While PL of the stem-QW was unchanged, PL of the arm-QW and the QWR changed markedly in shape and intensity with changes in gate bias $V_g$. Under strong negative bias $V_g\sim$-0.4 V, where electrons were depleted from the QWR, the PL of the QWR was quenched. This is because the strong electric field separates the photo-excited electrons and holes around the QWR, and drains some of the electrons down in the stem well. For weak negative bias $V_g\sim$-0.1 V, a peak denoted by X appeared. It is assigned as exciton PL because neutral excitons are expected to stay in the QWR. For zero and positive bias $V_g>0$ where electrons were accumulated in the QWR, the exciton peak X became weaker while another low-energy peak X${}^-$ assigned to charged excitons appeared and became stronger. At $V_g=0.1\sim$0.2 V, X$^-$ completely dominated the PL spectrum of the QWR. At high positive bias, PL of the QWR became asymmetric and red-shifted, which is assigned to a band-to-band optical recombination between an electron plasma and the minority holes. The present observation and the assignment of excitons, charged excitons, and band-to-band recombination are analogous to those reported in 2-D QWs \cite{Bar-JosephPRL,ShieldsPRB}, but here for the first time are achieved in a 1-D QWR of high optical quality.

\begin{figure}
\includegraphics[width=.3\textwidth]{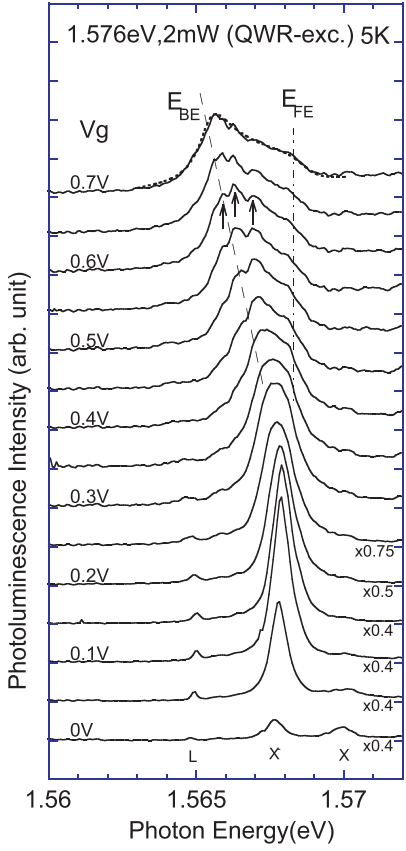}
\caption{
PL spectra of QWR at 5K for positive gate voltages $V_g$ measured in 0.05 V steps. 
The dashed line and a vertical dash-dot line are drawn to guide the eyes, and indicate the red-shifted band edge $E_{BE}$ and Fermi edge $E_{FE}$ singularity. Data at $V_g$=0.7V are fitted with a calculation shown by a dotted curve. 
}
\label{2}
\end{figure}

\begin{figure}
\includegraphics[width=.3\textwidth]{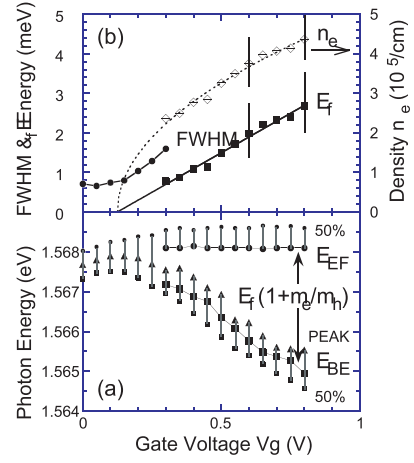}
\caption{
Plots of singular energy positions against gate voltage $V_g$. (a) Triangles, large squares, and large dots show peak energy, band edge $E_{BE}$, and Fermi edge $E_{FE}$, respectively, with error bars limited by half-maximum positions (small squares and dots). (b) Plots of FWHM energy of charged-exciton PL (dots), electron Fermi energy $E_f$ (squares), 1-D electron density $n_e$ (diamonds, right vertical axis), derived with $m_e/m_h=0.067/0.38$, $n_e=2k_f/\pi$, and $E_f=\hbar^2k_f^2/2m_e$. 
}
\label{3}
\end{figure}

To highlight the 1-D charged excitons and 1-D band-to-band recombination in the QWR, we show, in Fig. 2, PL spectra of the QWR with spectral resolution of 0.15 meV at 5 K for positive gate voltage $V_g$ measured in 0.05 V steps between 0 and 0.7 V. Here, we minimized carrier heating by reducing the incident photon energy to 1.576 eV, which directly created carriers only in the QWR with excess energy of 8 meV. The incident power was raised to 2 mW to compensate the reduced absorption cross section. The estimated electron-hole density $n_{eh}$ photo-excited in the QWR was  $\sim$10${}^4$ cm${}^{-1}$.  A reduction of excitation intensity by one order of magnitude resulted in almost no change in PL spectral shapes except for signal-to-noise ratios. 

At $V_g\sim$0 V, the two peaks of excitons X and charged excitons X${}^-$ are simultaneously observed. A very small sharp peak (L) at a lower energy (1.565 eV) is believed to be related to a localized state in the QWR, as it intermittently disappears as one moves along the wire. The energy separation between X and X${}^-$ is 2.3 meV. This is the first measurement of the extra binding energy of a 1-D charged exciton in a QWR. This 1-D value is much larger than the reported values of 0.9-1.2 meV for the charged to neutral exciton separation in 2-D GaAs QWs \cite{Bar-JosephPRL,ShieldsPRB}. 

At high positive bias, where the PL originates from a band-to-band optical recombination between an electron plasma and the minority holes, the PL spectra becomes asymmetric and significant red-shifts occur in the onset. In short, the results of Fig. 2 demonstrate two key issues, a strong BGR in the 1-D QWR and the Fermi-edge singularity pinned at the charged exciton energy. 

In preparation for the clarification of these issues, we made a model calculation to interpret whether the observed PL spectrum at $V_g=0.7$ V is indeed explained by a band-to-band optical recombination.  The top dashed curve in Fig. 2 is a numerical calculation to fits the observed PL. It is calculated as a function proportional to $\int \rho_{\Gamma}^j(\epsilon) f^e(\epsilon) f^h(\epsilon) \delta_{\gamma}(\epsilon+E_{BE}-\hbar\omega) d\epsilon$ \cite{AronSSC}, where $\epsilon$ is the transition energy relative to the band edge $E_{BE}$, $\rho_{\Gamma}^j$ is the 1-D joint density-of-states with broadening $\Gamma$, $f^e$ and $f^h$ are the Fermi distribution functions of electrons and holes, $\delta_{\gamma}$ is the broadening function with damping $\gamma$ and photon energy $\hbar\omega$. We assumed $\Gamma$=0.5 meV with $\rho_{\Gamma}^j=\sqrt{ (\sqrt{\epsilon^2+\Gamma^2}+\epsilon)/(\epsilon^2+\Gamma^2)}$, $\delta_{\gamma}$  to be the delta function ($\gamma$=0 meV), Fermi energy $E_f=2.8$ meV, $E_{BE}$=1.5653 eV, hole temperature $T_h$=20 K, and electron temperature $T_e$=3 K, where hole temperature higher than 5K is due to optical injection and lower electron temperature represents enhanced kink due to Fermi-edge singularity. The good fit supports our model. 
Accordingly, the two singular edges (onset and cutoff) at the lowest and highest energy in the asymmetric PL observed in the high $V_g$ region are ascribed to the band edge $E_{BE}$ and the Fermi edge $E_{FE}$, respectively. 

It is now clear that the band edge $E_{BE}$ is red-shifted from its original position at X${}^-$, as more electrons are accumulated when the bias $V_g$ is increased. This demonstrates the large BGR effect observed in our wire due to the Coulomb interaction between many electrons and few holes. On the other hand, for all electron densities the Fermi edge $E_{FE}$ is unshifted from the energy position of X${}^-$. We emphasize that $E_{BE}$ and $E_{FE}$ in the lowest electron density limit both converge to X${}^-$, and not X, and also not the fundamental band edge in single-particle picture, which is higher than X by the exciton binding energy. It is significant that the band-to-band plasma PL for high $V_g$ occurs entirely at energies lower than X${}^-$. 

To be more quantitative, we plot, in Fig. 3 (a), singular energy positions in PL against gate voltage $V_g$=0$\sim$0.8 V. Triangles show the peak positions. For $V_g$=0$\sim$0.3 V, small squares and dots show the left and right half-maximum positions. For $V_g$=0.3$\sim$0.8 V, large and small squares show the 80\%- and half-maximum positions in the left side of the peak, while large and small dots show the Fermi-edge shoulder and its right half-value position. We regard the position of the large squares and dots as the band edge $E_{BE}$ and the Fermi edge $E_{FE}$, respectively, with the experimental error-bars within the triangles, small squares, and dots. 
The energy separation $E_{FE} - E_{BE}$ is approximated as $E_f(1+m_e/m_h)$ , where $m_e$ and $m_h$ are effective mass of electrons and holes. We determine $E_f$ by assuming $m_e/m_h=0.067/0.38$. The 1-D electron density $n_e$ and $E_f$ ($E_f=\pi^2 \hbar^2 n_e^2/8m_e$) are plotted in Fig. 3 (b), where the extrapolated least-square-fit line to $E_f$ intersects the $V_g$-axis at 0.125 V.  This shows that the accumulation of electrons starts at around $V_g$=0.125 V, and proceeds with $E_f$ increasing linearly. At $V_g$=0.8 V, $E_f$ and $n_e$ reach 2.7$\pm$0.6 meV and 4.4$\pm$0.5$\times 10^5$cm${}^{-1}$, and $E_{BE}$ is red-shifted from X${}^-$ by 2.9$\pm$0.5 meV. The 1-D BGR red-shift is large compared  to the reduced Fermi energy. 
$E_f \times m_e/m_h$, the kinetic energy of a hole contributing to PL at the Fermi edge, is 0.48 meV, which is comparable to the lattice temperature (5 K). 

Figure 3 (b) also shows the FWHM of the charged-exciton PL (X${}^-$) for $V_g$ in the range 0-0.3 V. It is as sharp as 0.7 meV for low $ V_g$. The width starts to increase from $V_g$=0.15 V, and changes to two split peaks above $V_g$=0.3 V. 
The range of $V_g$=0.15$\sim$0.3 V corresponds to $n_e$=1.0$\sim$2.5$\times 10^5$ cm${}^{-1}$, a value close to recent predictions for the Mott density of excitons in GaAs QWRs \cite{DassarmaPRL}. In the present experiment, $n_e$ is not very high, so that electron Fermi energy is always much smaller than the estimated exciton binding energy of 14 meV. It is interesting that band-to-band plasma PL is observed in such low electron densities because it is expected to prevail only in the high electron density limit, where the kinetic energy dominates the Coulomb energy of the electron gas. Though the plasma PL picture seems to explain PL shapes in high $V_g$ region, we should aware that it is still strongly influenced by Coulomb interactions. 

We wish to comment here in Fig. 2 that the exciton peak (X) does not disappear completely even at the highest $V_g$, and remains unshifted at $\sim$2 meV above the Fermi edge $E_{FE}$, and that small unshifted ripples marked by arrows between $E_{FE}$ and $E_{BE}$ in PL spectra for $V_g>0.5$ V are currently not understood. 

\begin{figure}
\includegraphics[width=.4\textwidth]{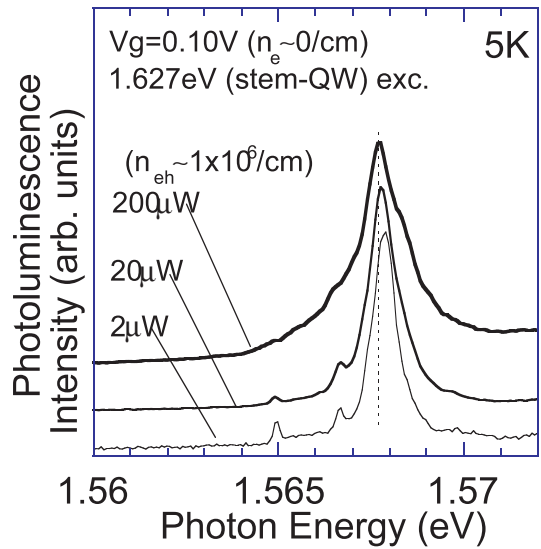}
\caption{
The excitation-density-dependent PL spectra of the QWR at 5K for $V_g$=0.1 V. Thin, medium, and thick curves show PL for excitation intensities of 2, 20, and 200 $\mu$W, respectively. 
}
\label{4}
\end{figure}

To explore differences in many-body effects between many-electron-few-hole systems and many-electron-many-hole systems in a QWR, we investigated the PL with higher photo-excited electron-hole density. For this purpose, the photon energy of excitation was raised to 1.627 eV to bring it in resonance with the exciton absorption of the stem-QW. The PL intensity and spectra obtained for weak (2 $\mu$W) photo-excitation were similar to spectra shown in Fig. 2.  Figure 4 shows PL spectra of the depleted ($n_e \rightarrow$0 cm${}^{-1}$) QWR at $V_g$=0.1 V.  Weak (2 $\mu$W), medium (20 $\mu$W), and strong (200 $\mu$W) photo-excitation intensities correspond to estimated electron and hole densities $n_{eh}$ of 10${}^4$, 10${}^5$, and 10${}^6$ cm${}^{-1}$, respectively. 

Note that increased photo-excitation density $n_{eh}$ results in only slight red-shifts, 0.2 meV even for $n_{eh}\sim$1$\times$10${}^6$ cm${}^{-1}$ (thick curve). This is in stark contrast to the significant BGR by a few meV caused by excess electron density $n_e$=2.5$\sim$4.4$\times$10${}^5$ cm${}^{-1}$ in Fig. 2.  
Note also the significantly broadened tails at high photoexcitation, while the top part of the PL peak was not broadened. The full widths of 90\%-maximum of three curves of different $n_{eh}$ were all 0.3 meV.
The persistency of this sharp peak against increased screening and state filling by electrons and holes in the neutral plasma shows the significant contribution of the Coulomb interaction. Thus, the minor red-shifts, or BGR, are probably the result of a cancellation of Coulomb interaction energy due to charge neutrality among equal densities of electrons and holes.

These neutral plasma results are consistent with previous results on QWRs showing broadening and no red-shift of PL in undoped QWRs under wide range of photo-excitation intensities \cite{WegscheiderPRL,AmbigapathyPRL}. We believe that lasing of QWRs at the fixed and same energy position as the excitonic PL \cite{WegscheiderPRL} could be related to the cancellation of Coulomb correlation in neutral photo-excited electron-hole plasmas. 

In summary, results from gated 1-D electron systems in a QWR of enhanced structural perfection indicate significant 1-D BGR in optical emission. The observation of 1-D charged exciton with large binding energy and its evolution to Fermi-edge singularity also highlight the crucial importance of Coulomb interaction effects in the quantum wire. The weak BGR observed in PL from photo-excited neutral electron-hole plasmas is likely to result from cancellation of Coulomb interaction effects.

We thank Drs. M. J. Matthews, R. dePicciotto, P. Platzman, and T. D. Harris for technical assistance and helpful discussion. 
One of the authors (H. A.) acknowledges the financial support from the Ministry of Education, Culture, Sports, Science and Technology, Japan.

\end{document}